\documentstyle[11pt]{article} 
\def\A{\it A}
\def\ds{\displaystyle}
\large{
\centerline{\bf One-dimensional relativistic dissipative system with constant }
\vskip1pc
\centerline{\bf force and its quantization}
}
\centerline{G. L\'opez$^{1}$, X. E. L\'opez$^{2}$ and H. Hern\'andez$^{1}$} 
\centerline{$^{1}$Departamento de F\'{\i}sica de la Universidad de Guadalajara}
\centerline{Apartado Postal 4-137}
\centerline{44410 Guadalajara, Jalisco, M\'exico}   
\centerline{$^2$ Facultad de Ciencias de la UNAM}
\centerline{Apartado postal 70-348, Coyoac\'an 04511 M\'exico D.F.} 
\centerline{PACS: 03.20.+i, 03.30.+p, 03.65.-w} 
\centerline{March, 2005}
\begin{document}  
\large{
\centerline{\bf ABSTRACT}
}
\vskip1cm
For a relativistic particle under a constant force and a linear velocity dissipation force, a
constant of motion is found. Problems are shown for getting the Hamiltonian of this system. Thus,
the quantization of this system is carried out through the constant of motion and using the
quantization on the velocity variable. The dissipative relativistic quantum bouncer is outlined within this
quantization approach.
\vfil\eject
\large{
\leftline{\bf 1. INTRODUCTION}    
}  
\vskip0.5pc
It is known that dynamical systems with dissipation (dissipative systems) represent some
difficulties for their right formulation in terms of Lagrangian and Hamiltonian formalisms
(L\'opez, 1996) . Normally this dissipation is included in the dynamical equations in a
phenomenological way  and through  a force which depends on the velocity of the particle. If the
Hamiltonian of a dissipative system is found, one proceeds to try to quantize the system. This has
been able to do for several nonrelativistic systems (Glauber and Man\'ko, 1984; Okubo, 1981;
L\'opez and Gonz\'alez, 2005; Mijatovic et al, 1985), but little is know about relativistic
dissipative systems. In this work, a constant of motion for a relativistic particle under a
constant force  and a linear dissipative force is obtained. The constant of motion at first order
in the dissipation parameter and the nonrelativistic limit of the constant of motion are analyzed,
and the problems for getting the Hamiltonian of the system are  outlined. Finally, the quantization
for the relativistic dissipative system, at first order on the dissipation parameter, is carried out
through the quantization of the velocity and the constant of motion.
\vskip2pc
\leftline{\bf 2. RELATIVISTIC CONSTANT OF MOTION}
\vskip1pc
The one-dimensional motion of a particle of mass $m$ at rest under a constant force, $f$, and a
linear dissipation force, $-\alpha v$, is governed by the equation 
$${d\over dt}\left({mv\over\sqrt{1-v^2/c^2}}\right)=-(f+\alpha v)\ ,\eqno(1)$$
where $v$ is the velocity of the particle, $\alpha$ is the dissipation parameter, and $c$ is the
speed of light. This equation can be written as the following autonomous dynamical system
$${d x\over dt}=v\eqno(2a)$$
$${d v\over dt}=-{f\over m}(1+\beta v)(1-v^2/c^2)^{3/2}\ ,\eqno(2b)$$ 
where $\beta$ is the constant defined as $\beta=\alpha/f$. A constant of motion for this system is
a function $K=K(x,v)$ which satisfies the following partial differential equation of first order
(L\'opez, 1999)
$$v~{\partial K\over\partial x}-{f\over m}(1+\beta v) (1-v^2/c^2)^{3/2}{\partial K\over\partial
v}=0\ .\eqno(3)$$
The general solution of this equation is given by
$$K_{\beta}(x,v)=G(\A(v)+fx)\ ,\eqno(4)$$
where $G$ is an arbitrary function, and $\A(v)$ has been defined as
$$\A(v)=m\int{v~dv\over(1+\beta v)(1-v^2/c^2)^{3/2}}\ .\eqno(5)$$
The result of the integration of (5) is
$$\A(v)=\cases{{\ds(1-\beta v)mc^2\over\ds\phi(v)}+{\ds m\beta c^3\over\ds(1-\beta^2c^2)^{3/2}}
\arcsin{\ds\beta c+v/c\over\ds 1+\beta v}& if~ $\beta c<1$\cr\cr\cr
{\ds m c v\over\ds\sqrt{ 1-v^2/c^2}}+{\ds m c^2(1-2 v/c-2v^2/c^2)
\over\ds 3(1+ v/c)\sqrt{1-v^2/c^2}}& if~$\beta c=1$\cr\cr\cr
{\ds(1-\beta v)mc^2\over\ds\phi(v)}+{\ds m\beta
c^3\over\ds(\beta^2c^2-1)^{3/2}}\log{\ds\psi(v)\over\ds 1+\beta v}& if~ $\beta c>1$}
\eqno(6)$$
where the functions $\phi(v)$ and $\psi(v)$ have been defined as
$$\phi(v)=\sqrt{1-v^2/c^2}~(1-\beta^2c^2)\eqno(7a)$$
and
$$\psi(v)=2\left(c^2\beta^2+\beta v+\sqrt{1-v^2/c^2}~\sqrt{c^2\beta^2-1}\right)\ .\eqno(7b)$$
The function $G$, appearing on (4), can be determinate through the criterion (L\'opez, 1996) of
having the usual relativistic energy expression for $\beta$ equal to zero (non dissipative case),
$$\lim_{\beta\to 0}K_{\beta}(x,v)={\ds mc^2\over\ds\sqrt{1-v^2/c^2}}+fx\ ,\eqno(8)$$
which brings about the result $G=I$ (the identity function)\footnote{one could add to (8) the
term
$-mc^2$ to have the usual energy expression for the nonrelativistic case}.
Therefore, the constant of motion for the system (2) can be chosen  as
$$K_{\beta}(x,v)=\A(v)+fx\ .\eqno(9)$$
This constant of motion brings about the damping effect on the trajectories in the phase space
$(x,v)$. Of course, due to multivalue functions of (6), the value of the constant of motion changes
for the trajectories going from the $x>0$ side to the $x<0$ side of this space, in order to get the
spiral falling down to the origin behavior of the trajectories. So, one may say that (9) represents
an "almost everywhere" constant of motion of the system (2), in the sense that the set of points
where these changes do occur has zero measurement (Hewitt and Stromberg, 1965), one one may call it
"local constant of motion."
\vskip1pc
For weak dissipation, one can also make a Taylor expansion on (5) of the term $(1+\beta v)^{-1}$ to
get the constant of motion of the following form
$$
K_{\beta}(x,v)={\ds mc^2\over\ds\sqrt{1-v^2/c^2}}
-m\beta\left[{vc^2\over\ds\sqrt{1-v^2/c^2}}-c^3\arcsin{v\over c}\right]+fx+\Phi(v)\ ,\eqno(10a)$$
where the function $\Phi$ is given by the expression
$$\Phi(v)=\sum_{n=2}^{\infty}(-1)^n\beta^n\left[-{c^2v^n\over(n-1)\sqrt{1-v^2/c^2}}
+{nc^2\over n-1}\int{v^{n-1}dv\over(1-v^2/c^2)^{3/2}}\right]\ .\eqno(10b)$$
Thus, at first order on the dissipation parameter $\beta$, one has 
$$K_{\beta}(x,v)=\gamma mc^2-m\beta c^3\biggl[{\gamma v\over c}-\arcsin{v\over c}\biggr]+fx\
,\eqno(11a)$$ where $\gamma$ has the usual expression,
$$\gamma={1\over\sqrt{1-v^2/c^2}}\ .\eqno(11b)$$
Note that the nonrelativistic limit ($v/c\ll 1$) must no be taken from the case $c\beta<1$ or the
case $c\beta=1$ but rather from the case $c\beta>1$. In this way, subtracting the term of energy
at rest, $mc^2$, on the case $c\beta>1$, the nonrelativistic constant of motion is given by
$$K_{\beta}(x,v)={m\over \beta^2}\biggl[\beta v-\log{(1+\beta v)}\biggr]+fx\ .\eqno(12)$$ 
Now, using the known expression (L\'opez and Hern\'andez, 1989; Kobussen,1979; Leubner, 1981) to
obtain the Lagrangian from the constant of motion,
$$L(x,v)=v\int {K(x,v)~dv\over v^2}\ ,\eqno(13)$$
one can get the Lagrangian and generalized linear momentum ($p=\partial L/\partial v$) given by
$$L_{\beta}(x,v)=B(v)-fx\eqno(14a)$$
and 
$$p(v)=C(v)\ ,\eqno(14b)$$
where the functions $B(v)$ and $C(v)$ are given in the appendix. in particular, the Lagrangian and the
generalized linear momentum for a relativistic particle with dissipation at first order in the dissipation
parameter $\beta$ (expression $11a$) are
$$L_{_1}(x,v)=-{mc^2\over\gamma}-m\beta c^3\biggl[{v\over 2c}\log{\gamma-1\over\gamma+1}+
{\arcsin{\ds v\over\ds c}\over\gamma}+{v\over c}\log{v\over c}\biggr]-fx\eqno(15a)$$
and
$$p(v)=\gamma mv+\beta mc^2\biggl[-{5\over 2}+{v\gamma\over c}\arcsin{v\over c}-{3\over 2}
\log{v\over c}\biggr]\ .\eqno(15b)$$
Similarly, for the nonrelativistic case with dissipation (expression (12)), one has 
$$L_{_2}(x,v)={m\over\beta^2}(\beta v-1)\log{(1+\beta v)}\ ,\eqno(16a)$$ 
and
$$ p(v)={m\over\beta}\biggl[\log{(1+\beta v)}-{1-\beta v\over 1+\beta v}\biggr]\ .\eqno(16b)$$
As one can see from (14b), ($A_2$), (15b) and (16b), it is not possible to have the inverse relation
$v=v(p)$. Therefore, their Hamiltonians are expressed only in an implicit way through the
constants of motion (9), (12) and (11).
Thus, the quantization of the system (1) can not be carried out with the standard Shr\"odinger
equation (Messiah, 1958),
$$i\hbar{\partial\Psi\over\partial t}=\widehat H(\widehat x,\widehat p)\Psi\ ,\eqno(17a)$$
or Heisenberg equation,
$$ i\hbar{d\widehat\xi\over dt}=[\widehat\xi,\widehat H]+i\hbar {\partial\widehat\xi\over\partial
t}\ ,\eqno(17b)$$
where $\widehat\xi$ is any time depending observable. The Feynman path quantization (Feynman and
Hibbs, 1965) is, in principle, possible to use here since one has gotten the Lagrangians (14a),
(15a) and (16a). However, the analytical functions appearing in these expression represent a real
challenge for the quantization with the path integration  method. One must also note from relations (14a) and 
(15a) that these can not be expressed in a covariant way since Lorentz transformations do  not leave invariant
this dissipation system.
\vskip3pc
\leftline{\bf 3. QUANTIZATION OF THE CONSTANT OF MOTION}
\vskip1pc
We are interested here in the quantization of the system (1) at first order in the dissipation 
parameter $\beta$, characterized by the constant of motion (11a). As it was mentioned above, the
quantization using the Hamiltonian or the Lagrangian approaches does not look plausible due to the
implicit form of the Hamiltonian and the complicated expression for the Lagrangian. However, one
can try to use the idea of quantizing the velocity (L\'opez, 2000) through the obvious
expression
$$\widehat v=-i{\hbar\over m}{\partial\over\partial x}\ .\eqno(18)$$
In this way, one can use directly the constant of motion of our system to make its quantization
through the equivalent Schr\"odinger equation
$$ i\hbar{\partial\Psi\over\partial t}=\widehat K(\widehat x, \widehat v)~\Psi\ ,\eqno(19)$$
where $\widehat K$ is the Hermitian linear operator associated to the constant of motion $K$ and
which has units of energy. 
\vskip1pc
Thus, let us consider the constant of motion (11a). From (19) and (11a), the equation obtained is
given by
$$i\hbar{\partial\Psi\over\partial t}=\left\{{mc^2\over\sqrt{1-{{\widehat v}^2/ c^2}}}
-m\beta c^3\biggl[{\widehat v\over\sqrt{1-{{\widehat v}^2/ c^2}}}-
\arcsin{\widehat v\over c}\biggr]+fx\right\}\Psi\ .\eqno(20)$$
Equation (20) represents an stationary problem, and the proposition
$$\Psi(x,t)=\psi(x)exp{\left(-i{ E~t\over\hbar}\right)}\eqno(21)$$
transforms (20) to an eigenvalue problem,
$$\left\{{mc^2\over\sqrt{1-{{\widehat v}^2/ c^2}}}
-m\beta c^3\biggl[{\widehat v\over\sqrt{1-{{\widehat v}^2/ c^2}}}-
\arcsin{\widehat v\over c}\biggr]+fx\right\}\psi=E\psi\ .\eqno(22)$$
One can see from this expression that it is better to look for its solution in the velocity
representation, which is given by applying the Fourier transformation to the function $\psi$,
$$\phi(v)={\it F}[\psi(x)]={1\over\sqrt{2\pi}}\int_{\it R}e^{i{ mvx/\hbar}}
~\psi(x)~dx\ ,\eqno(23)$$
where the variable $v$ represents the velocity of the particle. Applying this Fourier
transformation to (22), one gets
$$\left\{{mc^2\over\sqrt{1-{{v}^2/ c^2}}}
-m\beta c^3\biggl[{ v\over\sqrt{1-{{v}^2/ c^2}}}-
\arcsin{v\over c}\biggr]+i{\hbar f\over m}{\partial\over\partial v}\right\}\phi=E\phi\ .\eqno(24)$$
This equation has the following solution
$$\phi_E(v)={1\over\sqrt{2v_o}}~e^{i{\ds m^2c^3\over\ds\hbar f}
\biggl[\ds
\arcsin{\ds v\over\ds c}-2\beta c\sqrt{1-{v^2\over c^2}}-\beta v\arcsin{\ds v\over\ds c}-{\ds E
v\over \ds mc^3}\biggr]}\ ,\eqno(25)$$
where $v_o$ represent some maximum velocity of the particle ($v_o\le c$), and 
$[-v_o,v_o]$ is the intervale of velocities where the normalization of the function  (25)
has been carried out,
$$\int_{-v_o}^{v_o}|\phi_E(v)|^2dv=1\ .\eqno(26)$$
One also has that
$$\langle\phi_{E}|\phi_{E'}\rangle=\int\phi_E^*(v)\phi_{E'}(v)~dv={\hbar f\over m}\delta(E-E')\
.\eqno(27)$$
The spectrum of energies of the particle is continuous because of the form of the potential, $fx$,
and the general solution of (20) can be written, using (21) and (25), as
$$\Phi(v,t)=\int A(E)\phi_E(v)e^{-iEt/\hbar}~dE\ ,\eqno(28)$$
where the coefficient $A(E)$ is determinate by the initial condition $\Phi(v,0)$ in the following
way
$$A(E)={m\over \hbar f}\int \phi_E^*(v)\Phi(v,0)~dv\ .\eqno(29)$$
Now, in the case the potential be of the form
$$V(x)=\cases{fx&if~~$x>0$\cr\cr\infty & if~~$x\le 0$}\eqno(30)$$
which would correspond to the one-dimensional dissipative relativistic bouncer problem,
one requires that $\psi(0)=0$ and $\psi(x)=0$ if $x<0$~ for the solution of (22). This condition
brings about the discrete spectrum of the system and  can be written in the velocity
representation, using the inverse Fourier transformation, in the following way
$$0=\psi(0)={\it F}^{-1}[\phi_E(v)]\biggr|_{x=0}={1\over\sqrt{2\pi}}\int\phi_E(v)~dv\ .\eqno(31)$$
Given the set of eigenvalues, $\{E_n\}$, the eigenfunctions are given by
$$\phi_n(v)={1\over\sqrt{2v_o}}~e^{i{\ds m^2c^3\over\ds\hbar f}
\biggl[\ds
\arcsin{\ds v\over\ds c}-2\beta c\sqrt{1-{v^2\over c^2}}-\beta v\arcsin{\ds v\over\ds c}-{\ds E_n
v\over\ds  mc^3}\biggr]}\ ,\eqno(32)$$
and the general solution can be written as
$$\Phi(v,t)=\sum_nA_n\phi_n(v)~e^{-iE_nt/\hbar}\ ,\eqno(33)$$
where, using the orthogonality $\langle\phi_n|\phi_{n'}\rangle=\delta_{n,n'}$, the coefficients
$A_n's$ are determinate by the initial condition
$\Phi(v,0)$ and through the following expression
$$A_n=\int_{-v_o}^{v_o}\phi_n^*(v)\Phi(v,0)~dv\ .\eqno(34)$$
The wave function in the $x$ representation is gotten through the inverse Fourier transformation,
$$\psi_n(x)={1\over\sqrt{2\pi}}\int e^{-i{mvx/\hbar}}\phi_n(v)~dv\ .\eqno(35)$$
Note that the above quantization process can be done in general for the constant of motion (9) and
any case defined by (6).
\vfil\eject
\leftline{\bf CONCLUSION}
\vskip0.5pc
For a relativistic particle under a constant force and a dissipative force proportional to the
velocity, a local constant of motion has been found , and one has outlined the problem to get its
Lagrangian and Hamiltonian in general. To quantize this system , Schr\"odinger quantization
approach has been used with this local constant of motion at first order on the dissipation
parameter and using the velocity representation of the wave function. Using this quantization
approach, we have outlined the dissipative relativistic quantum bouncer problem.
\vskip3pc
\vfil\eject
\leftline{\bf APPENDIX}
\vskip1pc\noindent
The function $B(v)$ is given by
$$B(v)=\cases{
{\ds mc^2\over\ds 1-\beta^2c^2}\left[-\sqrt{1-{v^2\over c^2}}+
\beta v\log{ 1+\sqrt{1-{v^2\over c^2}}\over\ds v/c}\right]+{\ds m\beta c^3f_1(v)\over\ds
(1-\beta^2c^2)^{3/2}}&if ~$\beta c<1$\cr\cr\cr
-mcv\log{\left({\ds 2(1+\sqrt{1-v^2/c^2}~)\over\ds v/c}\right)}+{\ds mc^2 f_2(v)\over\ds 3} 
&if~$\beta c=1$\cr\cr\cr 
{\ds mc^2\over\ds 1-\beta^2c^2}\left[-\sqrt{1-{v^2\over c^2}}+
\beta v\log{ 1+\sqrt{1-{v^2\over c^2}}\over\ds v/c}\right]+{\ds m\beta c^3f_3(v)\over\ds
(1-\beta^2c^2)^{3/2}}&if~$\beta c>1$}
\eqno(A_1)$$
where $f_1$,$f_2$ and  $f_2$ are defined as
$$f_1(v)=v\int{\ds\arcsin{\left(\beta c+v/c\over 1+\beta v\right)}~dv\over v^2}\ ,
\eqno(\alpha_1)$$ 
$$f_2(v)={\ds -1-{v\over c}+{v^2\over c^2}+{3v\over c}\sqrt{1-{v^2\over c^2}}
\log{\left({\ds 2(1+\sqrt{1-v^2/c^2}~)\over v/c}\right)}\over \sqrt{1-v^2/c^2}}\eqno(\alpha_2)$$
and
$$f_3(v)=v\int{\ds\log{\left(\psi(v)\over 1+\beta v\right)}~dv\over\ds v^2} ,\eqno(\alpha_3)$$
where the function $\psi$ is given by (7b). The function $C(v)$ is given by
$$C(v)=\cases{
{\ds mc^2\over\ds 1-\beta^2c^2}\left[{\ds v-\beta c^2\over\ds c^2\sqrt{1-{v^2\over
c^2}}} +\beta\log{\left({1+\sqrt{1-v^2/c^2}\over v/c}\right)}\right]+
{\ds m\beta^3g_1(v)\over\ds (1-\beta^2c^2)^{3/2}}&if ~$\beta c<1$\cr\cr\cr
-mc\left[\log{\left({\ds 2(1+\sqrt{1-v^2/c^2}~)\over\ds v/c}\right)}-{\ds
1\over\ds\sqrt{1-v^2/c^2}}\right]+ {\ds mc g_2(v)\over 3} &if~$\beta c=1$\cr\cr\cr 
{\ds mc^2\over\ds 1-\beta^2c^2}\left[{\ds v-\beta c^2\over\ds c^2\sqrt{1-{v^2\over
c^2}}} +\beta\log{\left({1+\sqrt{1-v^2/c^2}\over v/c}\right)}\right]+
{\ds m\beta^3g_3(v)\over\ds (1-\beta^2c^2)^{3/2}}&if ~$\beta c>1$\cr}
\eqno(A_2)$$
where $g_1$, $g_2$ and $g_3$ are defined as
$$g_1(v)={f_1(v)\over v}+{\arcsin{\left({\ds\beta c+v/c\over\ds 1+\beta v}\right)}\over v}
,\eqno(\beta_1)$$
$$g_2(v)={g_{21}(v)+g_{22}(v)\over
(1-v^2/c^2)^{3/2}(1+\sqrt{1-v^2/c^2})}\ ,\eqno(\beta_2)$$
and
$$g_3(v)={f_3(v)\over v}+{1\over v}\log{\left({\psi(v)\over 1+\beta v}\right)}\ .\eqno(\beta_3)$$
The functions $g_{21}$ and $g_{22}$ have been defined as
$$g_{21}(v)=(-4-{v\over c}-{3v^2\over c^2}+{v^3\over c^3})(1+\sqrt{1-{v^2\over
c^2}})\eqno(\delta_1)$$ 
and
$$g_{22}(v)=
3(1-{v^2\over c^2})\left[-{v^2\over c^2}+1+\sqrt{1-{v^2\over c^2}}\right]
\log{\left({\ds 2(1+\sqrt{1-v^2/c^2})\over\ds v/c}\right)}\eqno(\delta_2)$$
\vfil\eject
\leftline{\bf REFERENCES}
\vskip0.5pc
\obeylines{Feynmann, R.P. and Hibbs, A.R.,(1965). {\it Quantum Mechanics} 
\hskip2cm {\it and Path Integrals}. McGraw-Hill, New York.
Glauber, R. and Man'ko, V.I. (1984). Sov. Phys JEPT {\bf 60},
\hskip2cm 450.
Gradshteyn, I.S. and Ryzhik, I.M. (1980). {\it Table of Integrals,
\hskip2cm Series, and Products} Academic Press, San Diego.
Hewitt, E. and Stromberg, K. (1965).{\it Real and Abstract 
\hskip2cm Analysis}, Springer-Verlag, New York.
Kobussen, J.A. (1979). Acta Phys. Austr. {\bf 51}, 193.
Leubner, C. (1981). Phys. Lett. A {\bf 86},2.
L\'opez, G. and Hern\'andez, J.I. (1989). {\it Hamiltonian and 
\hskip2cm Lagrangian for one-Dimensional Autonomous Systems}. 
\hskip2cm Ann. of Phys. {\bf 193},1.
L\'opez, G. (1996). {\it One-Dimensional Autonomous Systems and 
\hskip2cm Dissipative Systems}. Ann. of Phys. {\bf 235},2,372.
\L\'opez, G. (1999). {\it Partial Differential Equations of First 
\hskip2cm Order and Their Applications to Physics}, 
\hskip2cm World Scientific, Singapore-N. J.-London-Hong Kong.
L\'opez, G. (2000). {\it Quantization of a Constant of Motion for
\hskip2cm the Harmonic Oscillator with a Time-Explicitly 
\hskip2cm Depending Force}. quant-ph/0006091. 
L\'opez, G. et al (2001). {\it Quantization of the one-dimensional 
\hskip2cm particle motion with dissipation}. 
\hskip2cm Mod. Phys. Lett. B, {\bf 15}, 22, 965.
L\'opez, G. and Gonz\'alez, G. (2004). {\it Quantum Bouncer with 
\hskip2cm Dissipation}. Int. Jou. Theo. Phys. {\bf 43},10,1999.
Messiah, A. (1958). {\it Quantum Mechanics Vol. I}.
\hskip2cm John Wiley and Sons.
Mijatovic, M. et al (1985). Hadronic J. {\bf 7},5,1207.
Okubo, S. (1981). Phys. Rev. A, {\bf 23}, 2776.
}

\end{document}